\documentstyle[prl,aps,epsf,floats]{revtex} \def\narrowtext{} \tighten 
\twocolumn
\input epsf.sty

\begin{document}
\draft

\title{Quasiparticles and Energy Scaling in Bi$_2$Sr$_2$Ca$_{n-1}$Cu$_n$O$_{2n+4}$ ($\it{n}$=1-3): \\Angle-Resolved Photoemission Spectroscopy}

\author{
        T. Sato,$^1$
        H. Matsui,$^1$
        S. Nishina,$^1$
        T. Takahashi,$^1$
        T. Fujii,$^{2,*}$
        T. Watanabe,$^{3,\dagger}$
		and A. Matsuda$^{2,3}$
       }
\address{
         (1) Department of Physics, Tohoku University, 980-8578 Sendai, Japan\\
         (2) Department of Applied Physics, Faculty of Science, Science University of Tokyo, Tokyo 162-8601, Japan\\
         (3) NTT Basic Research Laboratories, Atsugi 243-0198, Japan\\
         }
\address{%
\begin{minipage}[t]{6.0in}
\begin{abstract}
		Angle-resolved photoemission spectroscopy (ARPES) has been performed on the single- to triple-layered Bi-family
		high-{\it T$_c$} superconductors (Bi$_2$Sr$_2$Ca$_{n-1}$Cu$_n$O$_{2n+4}$, $\it{n}$=1-3).  We found a sharp quasiparticle peak as well as a pseudogap
		at the Fermi level in the triple-layered compound.  Comparison among three compounds has revealed a universal rule
		that the characteristic energies of superconducting and pseudogap behaviors are scaled with the maximum {\it T$_c$}.
\typeout{polish abstract}
\end{abstract}
\pacs{PACS numbers: 74.72.Hs, 74.25.Jb, 71.18.+y, 79.60.Bm}
\end{minipage}}
\maketitle
\narrowtext
    	Over a decade, angle-resolved photoemission spectroscopy (ARPES) has played an important role in 
		developing our understanding of high-{\it T$_c$} superconductors (HTSCs).  A recent remarkable progress in 
		energy and momentum resolutions of ARPES enables a direct observation of ``quasiparticles" (QPs) in 
		HTSCs  which dominate the low-energy physics as well as the superconductivity. 
		The superfluid density \cite{Feng,HongZD} is discussed in terms of the temperature- and doping-dependence of the weight of QP peak. 
		ARPES lineshape at ($\pi$, 0) and a characteristic kink in dispersion have been interpreted as an interaction of 
		electrons with ($\pi$, $\pi$) magnetic mode \cite{JC,Adam1} and that with longitudinal optical (LO) phonon\cite{Lanzara}.  The lifetime of
		QPs deduced from ARPES has been found to have a direct relation to the optical conductivity\cite{Adam2} and the electrical
		resistivity \cite{Valla} and has been discussed in relation to the quantum critical behavior\cite{Valla}.  However, it is unknown at present 
		whether these characteristic features of QPs are universal to all HTSCs, since a majority of ARPES studies have been
		done on Bi$_2$Sr$_2$CaCu$_2$O$_{8+\delta}$ (Bi2212) which has CuO$_2$ bilayers.  It is thus important to perform a systematic ARPES study 
		on HTSCs with different numbers of CuO$_2$ layers in a unit cell to find universal properties in HTSCs.  Although Bi-family
		HTSCs (Bi$_2$Sr$_2$Ca$_{n-1}$Cu$_n$O$_{2n+4}$, $\it{n}$=1-3) look the most suitable to meet this aim, triple-layered Bi$_2$Sr$_2$Ca$_2$Cu$_3$O$_{10+\delta}$ (Bi2223)
		has been scarcely investigated by ARPES \cite{Janoviz} because of difficulty in growing a high-quality single crystal.
		
	In this Letter, we report a systematic high-resolution ARPES study on single- (Bi2201), double- (Bi2212), and triple- (Bi2223) 
	layered Bi-family HTSCs to investigate the universality of QP features as well as to find a possible scaling rule in physical properties.
	By performing ARPES on Bi2223 and comparing the result with the data of Bi2201 and Bi2212, we now address several important
	issues about (1) appearance of QP peak, (2) interaction of electrons with modes, (3) origin of pseudogap, and (4) energy scale
	which dominates the superconducting and pseudogap properties.

			High-quality single crystals of Bi$_2$Sr$_2$Ca$_2$Cu$_3$O$_{10+\delta}$ ({\it T$_c$}=108 K, nearly optimally doping) were grown by the 
			traveling solvent floating zone (TSFZ) method with a very slow growth rate and a steep temperature gradient\cite{crystal}.
			The oxygen content $\delta$ was controlled by the temperature of post-annealing and/or the oxygen partial pressure. 
			The sharp x-ray diffraction pattern and the steep superconducting transition ($\Delta$$\it{T}<$3 K) in the DC susceptibility confirm 
			the high quality of crystal.  The typical size of crystal for ARPES measurement is 4 $\times$ 2 $\times$ 0.1 mm$^3$.   
			ARPES measurements were performed using a SCIENTA SES-200 spectrometer with a high-flux discharge lamp and 
			a toroidal grating monochromator.  We used the He I$\alpha$ (21.218 eV) resonance line to excite photoelectrons.  
			The energy and angular (momentum) resolutions were set at 9 meV and 0.2 deg (0.005\AA$^{-1}$), respectively.  
			Clean surface for ARPES measurement was obtained by $\it{insitu}$ cleaving in an ultrahigh vacuum of 4$\times$10$^{-11}$ Torr. 
			The Fermi level ({\it E}$_{F}$) of the sample was referenced to a gold film evaporated onto the sample substrate.  
			ARPES spectra of Bi2201 and Bi2212 were obtained with the same ARPES spectrometer under essentially the 
			same experimental conditions\cite{Sato}.

\begin{figure}[!t]
\epsfxsize=3.4in
\epsfbox{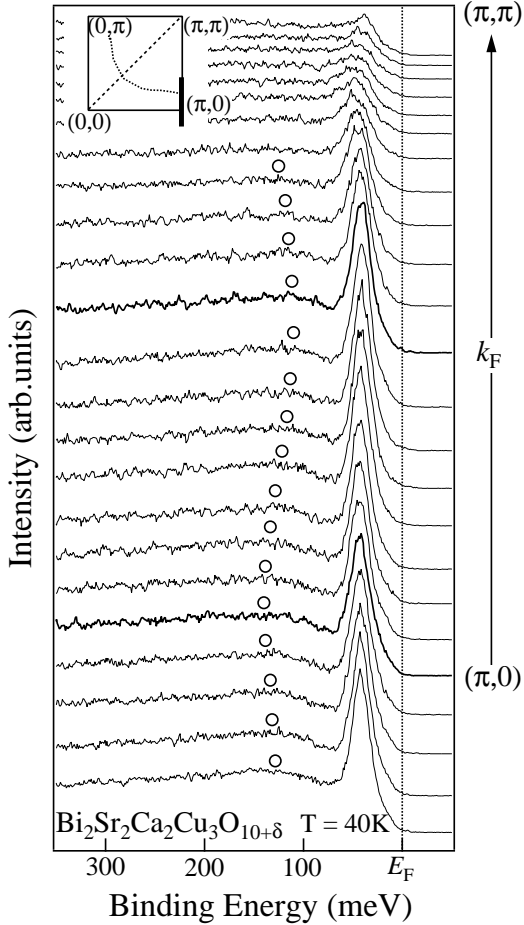}
\vspace{-0.2cm}
\caption{
ARPES spectra near {\it E}$_{F}$ of Bi$_2$Sr$_2$Ca$_2$Cu$_3$O$_{10+\delta}$ ({\it T$_c$}=108 K) measured at 40 K  along ($\pi$, 0)-($\pi$, $\pi$) cut.
  The Fermi vector {\it k}$_{F}$ = ($\pi$, 0.12 $\pi$) has been determined by the minimum-gap-locus method[11].
  Energy position of hump structure is denoted by open circles.
}
\label{fig1}
\end{figure}

			Figure 1 shows ARPES spectra of Bi2223 ({\it T$_c$}=108 K) measured at 40 K along ($\pi$, 0)-($\pi$, $\pi$) cut
			in the Brillouin zone.  We clearly find an almost non-dispersive sharp peak at 40-50 meV assigned 
			to the superconducting quasiparticle (QP) peak, together with a broad hump structure at 110-150 meV
			in the spectra.  It is thus evident that the ``hump-dip-peak" structure, observed so far in only bilayered HTSCs 
			Bi2212\cite{Dessau} and YBa$_2$Cu$_3$O$_{7-\delta}$ (Y123)\cite{Liu}, exists also in a triple-layered HTSC Bi2223.  As found in Fig. 1, 
			the hump structure shows a remarkable systematic dispersion and the QP peak looks to follow it although 
			the magnitude is considerably weak.  Similar close correlation between the dispersion of QP peakÊand hump
			is also seen in Bi2212\cite{Adam1}.   However, there are some quantitative differences between Bi2223 and Bi2212.  
			Interestingly, the hump-dip-peak structure is clearly resolved even at {\it k}$_{F}$ and the hump structure disperses back
			to the high binding energy after passing {\it k}$_{F}$ in Bi2223, while these features are not clearly seen in optimally-doped Bi2212 \cite{JC,Adam1}.
			In a simple model of electrons interacting with modes, a stronger coupling causes a stronger frequency (energy) dependence
			of electron self-energy\cite{Parks}.  In other words, the broad incoherent feature in ARPES spectrum (namely the hump) gains much
			intensity at the stronger interaction (coupling constant).  Therefore, the present experimental result suggests a stronger
			interaction of electrons with modes in triple-layered Bi2223 than in bilayered Bi2212.  At present there is no clear explanation 
			for this layer-number dependence of interaction strength.  A theoretical model to describe the relation of the coupling strength
			and the number of CuO$_2$ layers is necessary.

		          Figure 2 shows comparison of ARPES spectra in superconducting state measured at ($\pi$, 0) among three 
				  Bi-HTSCs at/near optimal doping.  The overall lineshape of Bi2212 and Bi2223 looks almost identical 
				  to each other in the energy position of hump and QP peak, as well as an intrinsic width of QP peak (14-16 meV)\cite{RefSF},
				  while that of Bi2201 is apparently different from the others, showing no sharp QP peak nor the hump-dip-peak structure,
				  although superconducting gap actually opens (leading edge clearly shifts to high binding energy).  This anomalous behavior,
				  the absence of a sharp QP peak in the single-layered Bi2201, may be explained in terms of a possible small $\it{c}$-axis 
				  superfluid density in Bi2201.  However, the difference in the superfluid density between Bi2201 and Bi2212 observed by
				  the microwave experiment \cite{Gaifullin} seems too small to account for the difference in the ARPES spectral lineshape, suggesting
				  the other mechanism to suppress the QP peak.
				  
				  			\begin{figure}[!t]
\epsfxsize=3.4in
\epsfbox{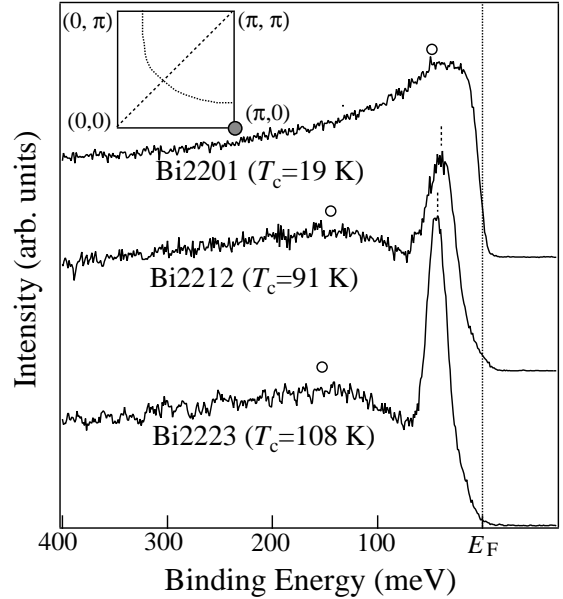}
\vspace{-0.2cm}
\caption{
Comparison of ARPES spectra at the superconducting state of Bi2201 ({\it T$_c$}=19 K),
 Bi2212 ({\it T$_c$}=91 K), and Bi2223 ({\it T$_c$}=108 K) measured at ($\pi$, 0) point. Measurement temperatures are
 13.5 K for Bi2201 and 40 K for both Bi2212 and Bi2223.  Energy positions of hump and quasiparticle
 peak are denoted by open circles and dashed lines, respectively.
}
\label{fig2}
\end{figure}

\begin{figure*}[!t]
\epsfxsize=6.8in
\epsfbox{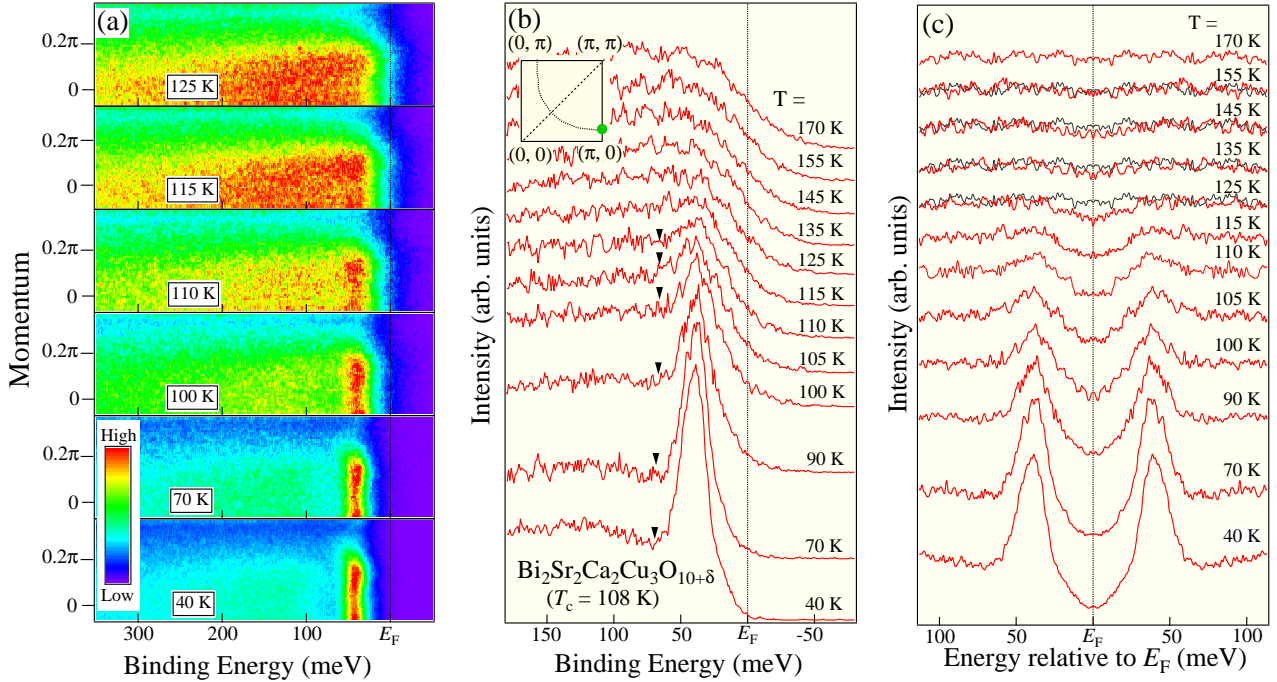}
\vspace{0.2cm}
\caption{
(a) Temperature dependence of ARPES intensity along ($\pi$, 0)-($\pi$, $\pi$) cut in Bi2223.  Vertical axis corresponds to the momentum
 along ($\pi$, 0)-($\pi$, $\pi$) cut while the abscissa shows the binding energy relative to {\it E}$_{F}$. Intensity is normalized to the peak maximum 
 at each temperature.
(b) Temperature dependence of ARPES spectra of Bi2223 at ($\pi$, 0)-($\pi$, $\pi$) crossing.  Intensity of spectra is normalized to the area 
under the curve.  The energy position of spectral break is indicated by arrows.
(c) Symmetrized ARPES spectra of Bi2223 at ($\pi$, 0)-($\pi$, $\pi$) crossing.  The 170-K spectrum (black line) is superimposed on each spectrum
 for comparison.
}
\label{fig3}
\end{figure*}

			 Next we discuss the temperature dependence of QP peak, superconducting gap and pseudogap.  
			 ARPES intensities of Bi2223 ({\it T$_c$}=108 K) along ($\pi$, 0)-($\pi$, $\pi$) cut at $\it{T}$= 40 - 125 K are plotted in Fig. 3(a). 
			 A pronounced narrow structure at 40 K (Fig. 3(a) bottom) is assigned to the QP peak.  With increasing temperature, 
			 the QP peak gradually reduces its intensity without changing its position, and finally vanishes around 115 K.  
			 This is more clearly seen in the ARPES spectra at {\it k}$_{F}$ (Fig. 3(b)), where a spectral break (indicated by triangles) 
			 seen in the 110-K spectrum disappears in the 115-K spectrum.  The survival of QP peak at temperatures slightly
			 above {\it T$_c$} is understood in terms of superconducting fluctuation, as in Bi2212 ({\it T$_c$}=91 K) where the QP peak at ($\pi$, 0) 
			 vanishes around 103 K\cite{Fedorov}.  We thus conclude that the temperature-dependent evolution of QPs is generic to double- 
			 and triple-layered HTSCs.

			A definite evidence for opening of a pseudogap above {\it T$_c$} in Bi2223 is demonstrated in Fig. 3(c),
			where ARPES spectra at ($\pi$, 0)-($\pi$, $\pi$) crossing (Fig. 3(b)) are symmetrized with respect to {\it E}$_{F}$\cite{Norman}. 
			We find that the superconducting gap smoothly evolves into the pseudogap above {\it T$_c$} (108 K).  
			It is also found that the energy scale of pseudogap is same as that of superconducting gap, and the pseudogap 
			gradually ``fills in" rather than ``closes" with increasing temperature.  These behaviors are similar to that observed
			in Bi2212 \cite{Fedorov,Loeser,HongNature}, indicating that pseudogaps in Bi2212 and Bi2223 have a same origin, most likely 
			a precursor pairing.
			In order to estimate $\it{T}$* at which the pseudogap closes, we have superimposed the 170-K spectrum on the 125-155 K spectra.
			We find that the spectra at 125-145 K clearly show a deviation from the 170-K spectrum near {\it E}$_{F}$, while the spectrum at 
			155 K appears to coincide well with that at 170 K.  This suggests that the pseudogap closes around $\it{T}$* $\sim$150 K, consistent 
			with the in-plane resistivity which shows a deviation from the $\it{T}$-linear dependence around 140-190 K\cite{transport}.

				In Fig. 4, we show the size of superconducting gap $\Delta$$_{max}$, the binding energy of hump at ($\pi$, 0),
				and $\it{T}$* as a function of maximum {\it T$_c$}  ({\it T$_c$}  at optimal doping) or the number of CuO$_2$ layers in a unit cell.
				The hump energy is regarded as a measure of a large pseudogap \cite{JC} and the values of $\it{T}$* are from the
				ARPES and the in-plane-resistivity measurements\cite{transport,Chong,Watanabe,Sato2}.   We find that all the energies are well on a straight
				line with respect to {\it T$_c$}, but not to the number of CuO$_2$ layers. This clearly indicates that the key superconducting
				and pseudogap properties of HTSCs are well scaled with the maximum {\it T$_c$}.  In other words, maximum {\it T$_c$} essentially
				controls key important superconducting and pseudogap properties and it is universal for Bi-family high-{\it T$_c$} superconductors.
				Furthermore, the scaling between the superconducting gap and pseudogap energy suggests that both are 
				closely related.
				
\begin{figure}[!t]
\epsfxsize=3.4in
\epsfbox{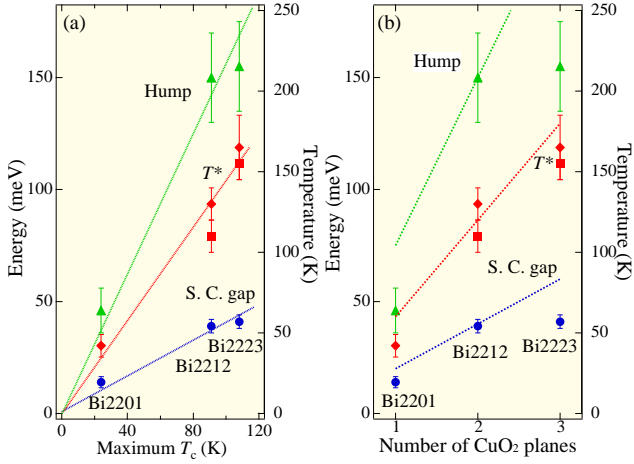}
\vspace{0.2cm}
\caption{
(a) Size of superconducting gap $\Delta$$_{max}$, the binding energy of hump at ($\pi$, 0), and $\it{T}$* as a function of maximum {\it T$_c$}.  Straight lines are the guides for eyes.
(b) Same as a function of the number of CuO$_2$ layers in a unit cell.
}
\label{fig4}
\end{figure}

				We now discuss these implications.  NMR study by Julien {\it et al.} has suggested a higher value of effective super 
				exchange interaction $\it{J}_{eff}$ in triple-layered HgBa$_2$Ca$_2$Cu$_3$O$_{8+\delta}$ (HBCO, {\it T$_c$}=121 K) than in bilayer Y123, from their analysis 
				of hyperfine coupling constant $\it{B}$\cite{Julien}.  Recent scanning tunneling spectroscopy and Raman spectroscopy have shown that
				the ratio of superconducting gap as well as $\it{T}$* between single-layered La$_{2-x}$Sr$_x$CuO$_4$ (LSCO) and bilayer Bi2212 is same
				as that of maximum {\it T$_c$} (scaling factor 91/38$\sim$2.4), and the scaling factor shows a striking agreement with the ratio of $\it{J}_{eff}$
				(measured by uniform magnetic susceptability) between LSCO and Bi2212\cite{Matsuzaki}.  Considering these experimental results 
				together with the scaling behavior found in the present ARPES study, we conclude that the key superconducting and pseudogap
				properties are closely related to the antiferromagnetism in HTSCs and the superconductivity is most likely due to spins in
				CuO$_2$ layers.
			
	     In conclusion, we have performed angle-resolved photoemission spectroscopy of triple-layered high-{\it T$_c$} 
		 superconductor Bi$_2$Sr$_2$Ca$_2$Cu$_3$O$_{10+\delta}$ ({\it T$_c$}=108 K).  We have found (1) sharp quasiparticle peak which disappears
		 slightly above {\it T$_c$} (115 K),  (2) hump-dip-peak structure at ($\pi$, 0) in superconducting state, and (3) pseudogap above
		 {\it T$_c$} ($\it{T}$*=150 K) which smoothly evolves into superconducting gap.  By a comprehensive comparison of Bi-family
		 high-{\it T$_c$} superconductors (Bi$_2$Sr$_2$Ca$_{n-1}$Cu$_n$O$_{2n+4}$, $\it{n}$=1-3), we found a universal rule that key energies which
		 dominate superconducting- and pseudo-gap properties are well scaled wtih maximum {\it T$_c$}.\\

			We thank M. R. Norman for useful discussions.  This work was supported by a grant from the MEXT of Japan.
			T.S. thanks the Sasakawa Scientific Research Grant from the Japan Science Society and the Japan Society 
			for Promotion of Science for financial support.\\

$^*$Present address: Department of Applied Physics, Waseda University, Tokyo 169-8555, Japan\\
$^{\dagger}$Present address: NTT Photonics Laboratories, Atsugi 243-0198, Japan\\

\end{document}